\pdfoutput=1
\documentclass[a4paper,fleqn,usenatbib]{mnras}

\usepackage{newtxtext,newtxmath,listings}

\usepackage[T1]{fontenc}
\usepackage{ae,aecompl}

\usepackage{graphicx}	
\usepackage{amsmath}	
\usepackage{amssymb}	

\title[Resolving merged doublet lines]{A simple non-parametric method for resolving merged doublet lines: Insights into complex kinematics and outflows}

\author[Villforth et al.]{
Carolin Villforth,\thanks{c.villforth@bath.ac.uk}
Tom L. Underwood,
Mark Tolson
and Nikhil Modha
\\
University of Bath, Department of Physics, Claverton Down, Bath, BA27AY, UK\\
}

\date{Accepted XXX. Received YYY; in original form ZZZ}

\pubyear{2018}

\begin{document}
\label{firstpage}
\pagerange{\pageref{firstpage}--\pageref{lastpage}}
\maketitle

\begin{abstract}
Doublet line emission and absorption is common in astronomical sources (e.g. [OIII], [OII], NaD, MgII). In many cases, complex kinematics in the emitting source can cause the 
doublet lines to merge, making characterisation of the source kinematics challenging. Here, we present a non-parametric method for resolving merged doublet emission when the line ratio and wavelength difference is known. The method takes as input only the line ratio and wavelength difference, using these quantities to resolve the components of the doublet without resorting to fitting (e.g. using multiple Gaussians) or making any assumptions about the components' line profiles (save that they are the same for both components). The method is simple, fast and robust.
It is also ideal for visualisation.
We show that the method recovers line profiles of merged emission lines in simulated data. We also show, using simulated data and mathematical analysis,
that the method does not significantly increase noise levels in the extracted lines, and is robust to background contamination. We demonstrate the strength of the method by applying it to strongly merged [OIII] 5007/4959~{\AA} in Active Galactic Nuclei (AGN). A \textsc{python} implementation of the method is provided in the appendix.
\end{abstract}




\section{Introduction}

Complex kinematics are common in astronomical objects \citep[e.g.][]{mullaney_narrow-line_2013, zakamska_quasar_2014, zakamska_discovery_2016}, with extremely asymmetric and 
non-Gaussian profiles ubiquitous. A non-parametric classification of such kinematics from emission or absorption lines using, for example, skew, kurtosis or percentiles can be used to quantify the kinematic structure of the emitting gas.

In many cases, the analysis of emission or absorption lines is complicated by the fact that many common atomic
lines are observed as doublets, e.g. [OIII] 5007/4959~{\AA}, 
[OII] 3727/3730~{\AA}, MgII 2796/2803~{\AA}, NaD 5892/5898~{\AA}, CaH/K 3935/3970~{\AA}. High velocities can cause the line profiles to merge \citep[see e.g.][]{greene_spectacular_2012, zakamska_quasar_2014,zakamska_discovery_2016}. And while simple kinematic structures can be fit using Gaussian mixture models \citep[e.g.][]{mullaney_narrow-line_2013}, 
reconstructing extremely complex line profiles requires a large number of free parameters. Moreover without careful fitting procedures, continuum emission can throw off fits using multiple components, especially in noisy data. Fitting can therefore be challenging when applied to large samples. 

A non-parametric approach to decompose doublet emission is therefore valuable for recovering the underlying line profiles for further analysis. 
Methods for decomposing doublet line profiles by exploiting symmetry in the different lines are known in the literature \citep[e.g.][]{junkkarinen_spectrophotometry_1983}. However no general mathematical framework for this approach is known.
\begin{figure*}
	\includegraphics[width=\columnwidth]{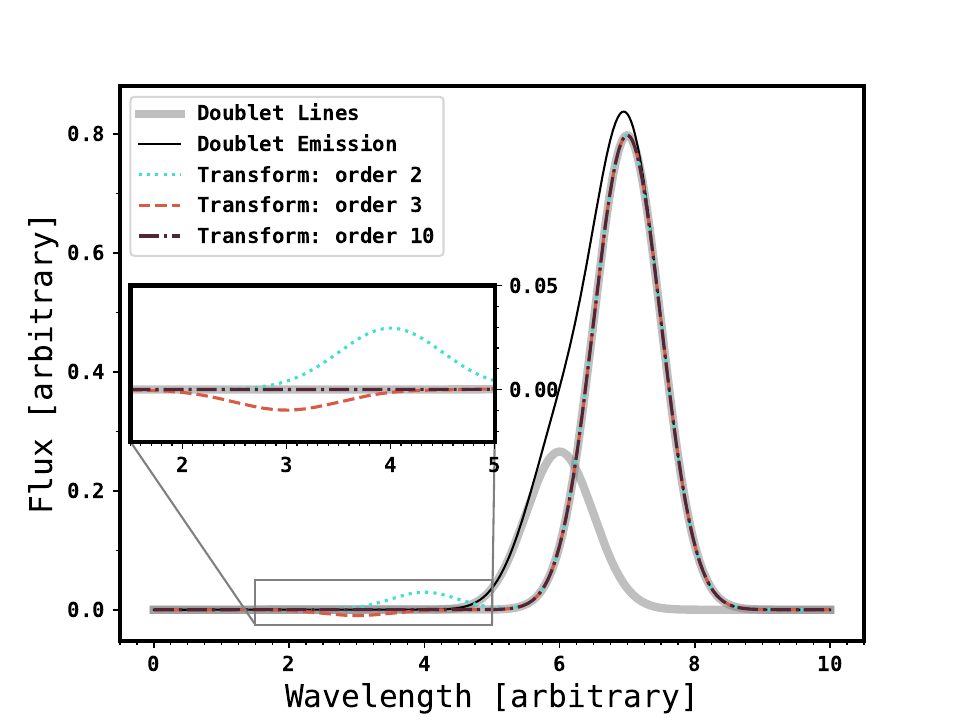}
    \includegraphics[width=\columnwidth]{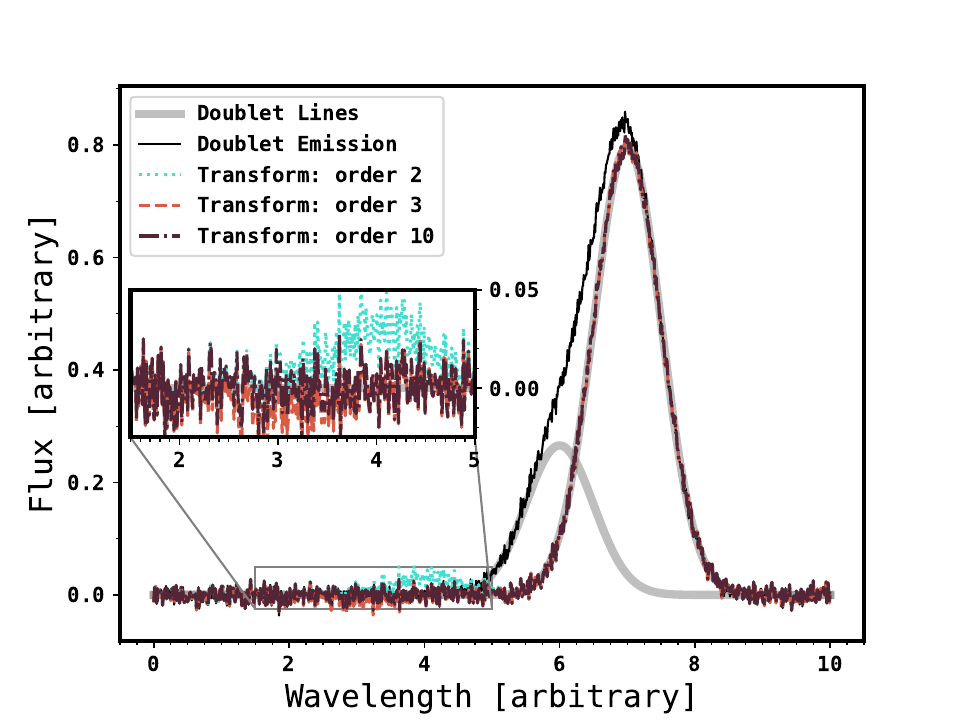}
    \caption{Simulations showing the performance of the method. Left: Application of the method to noise-free merged Gaussian emission lines for different numbers of iterations (order of the transform). Individual lines are shown in grey, and the fully-merged emission line is shown in dashed grey. The transformed data from orders 2, 3, and 10 are shown as cyan, orange and dark red dashed lines, respectively. The transform is well converged to the underlying line profile at order 10. Right: same as upper left, but with noise added to data. In both cases, the line ratio of the simulated doublet is $R=1/3$, the separation between the lines is $\delta=1$ and the width of each line is $\sigma=0.5$. The noise added in the right panel is Gaussian with $\sigma_{\text{noise}}=0.01$.}
    \label{fig:simu_Gaussian}
\end{figure*}

In this paper we focus on the general problem of merged doublet lines, and introduce and test a simple non-parametric method for resolving merged doublet line profiles when the line ratio and wavelength difference is known.  The method is a fast, simple alternative to parametric fitting techniques and requires as an input only the line ratio and wavelength separation. No estimate of the line shape is required. The method is introduced in Section~\ref{S:method}; as an example application, the method is applied to strongly-merged [OIII] 5007/4959~{\AA} doublets in Active Galactic Nuclei (AGN) in Section~\ref{S:results} and compared to a commonly used fitting approach; we discuss practicalities associated with using the method in Section~\ref{S:considerations}; and a summary of our findings, as well as possible use cases, and caveats and limitations of the method, are provided in Section~\ref{S:discussion}. An implementation of the method in \textsc{Python} is given in Appendix~\ref{python}, and discussion and derivations supplementary to the main text are given in Appendix~\ref{S:more_theory}.

\section{Method}
\label{S:method} 

Here, we describe the method used for doublet decomposition. The basic idea is that, since the line profiles are identical, one of the lines can be removed by subtracting the profile of the other. The method shifts the full line profile, multiplied by the line ratio $R$, by the wavelength difference $\delta$ and subtracts it from the data. This is repeated for a number of iterations. The continued iterations effectively correct for the fact that the full line profile, rather than an individual line profile, is used for subtraction.  We will show below that this simple procedure of shifting and subtracting the data recovers the underlying line profile correctly. A visualization of the method is shown in Fig.~\ref{fig:simu_Gaussian} (left panel).

We emphasise that the method makes no assumption regarding the underlying line profile, and hence is valid for any line profile. Furthermore, we will show later in Section~\ref{S:background} that continuum emission which can be approximated as linear across the wavelength range of interest does not affect the performance of the method. Emission lines not associated with the doublet, however, should be removed before the method is applied. Nevertheless we have examined the effect of line contamination, and discuss this topic later in Section~\ref{S:linecontamination}.

We will now explain the method in detail. Let $y(x)$ denote the flux observed at wavelength $x$ (i.e. the signal). This can be expressed as
\begin{equation}\label{I_defn}
y(x)=f_1(x)+f_2(x),
\end{equation}
where $f_1(x)$ and $f_2(x)$ are the two lines which make up the doublet, and we have ignored contributions such as noise and interference from the tails of 
other lines near the doublet -- we discuss such complications later.
The assumption which underpins our method is that the two lines which make up the doublet are \emph{similar}
in the geometrical sense, i.e. line 1 can be obtained from line 2 by rescaling the former, then shifting
the rescaled line along the $x$-axis by some amount. This can be expressed mathematically as follows:
the two lines, $f_1(x)$ and $f_2(x)$, are defined to be similar if there exists an $R$ and $\delta$ such that
\begin{equation}\label{same_shape_defn}
f_2(x)=Rf_1(x+\delta).
\end{equation}
Note that $R$ is the line ratio of line 2 relative to line 1, and $\delta$ is the $x$ position of peak 1 relative
to peak 2 (i.e. the wavelength difference between the two lines); if peak 1 is centred at $x=0$, then peak 2 will be centred at $x=-\delta$.
In other words $R$ is the line ratio and $\delta$ is the wavelength difference of the two peaks which make
up the doublet.
\footnote{Note that that $R$ and $\delta$ reflect the intrinsic properties of the two lines, and not their
manifestations in the doublet $y(x)$. E.g. if the doublet $y(x)$ exhibits two local maxima, then $R$ does
not necessarily correspond to the relative heights of these maxima, nor does $\delta$ necessarily correspond 
to their separation.}

\begin{figure*}
	\includegraphics[width=\columnwidth]{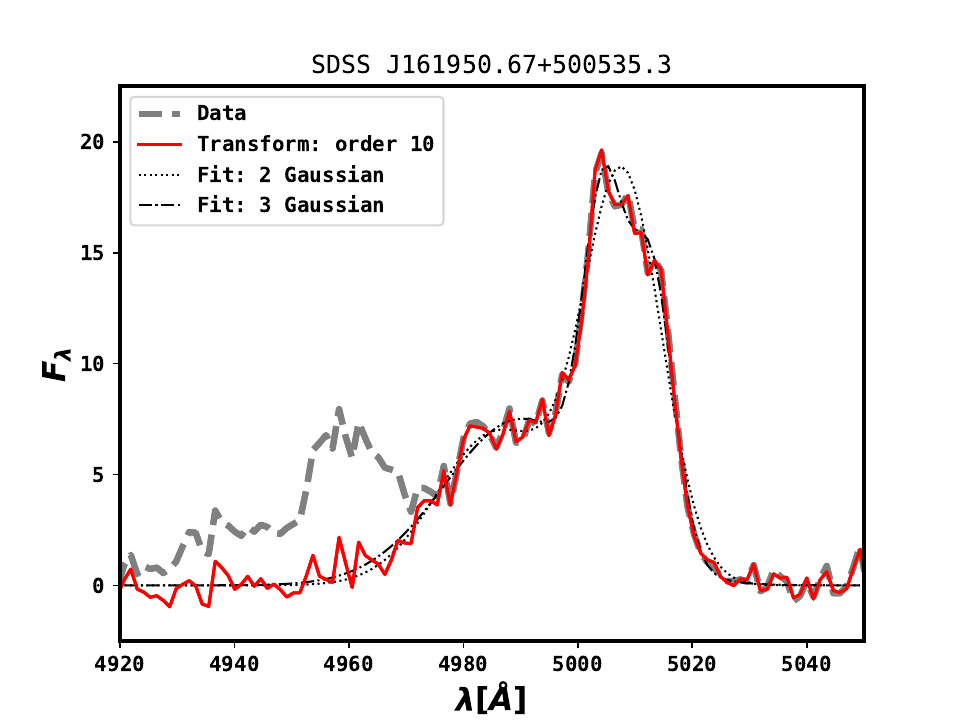}
    \includegraphics[width=\columnwidth]{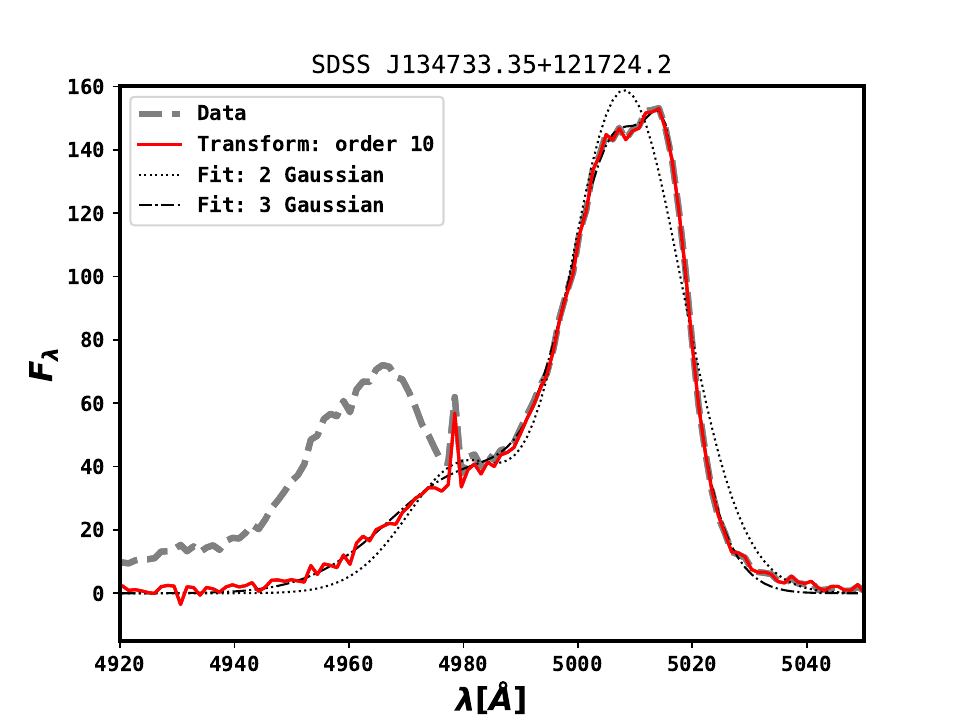}
    \includegraphics[width=\columnwidth]{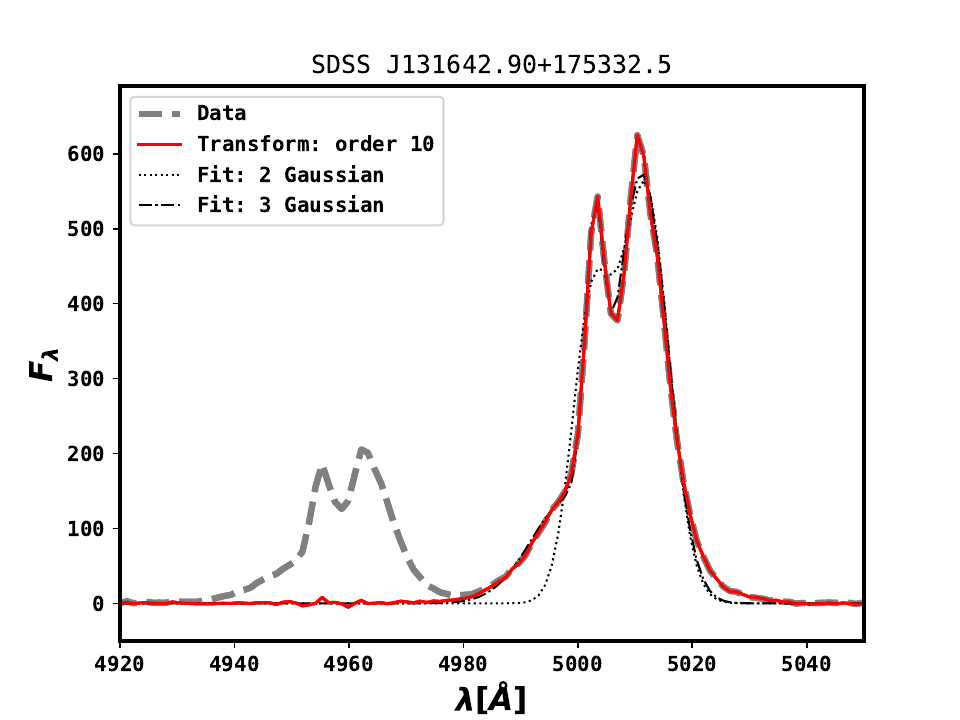}
    \includegraphics[width=\columnwidth]{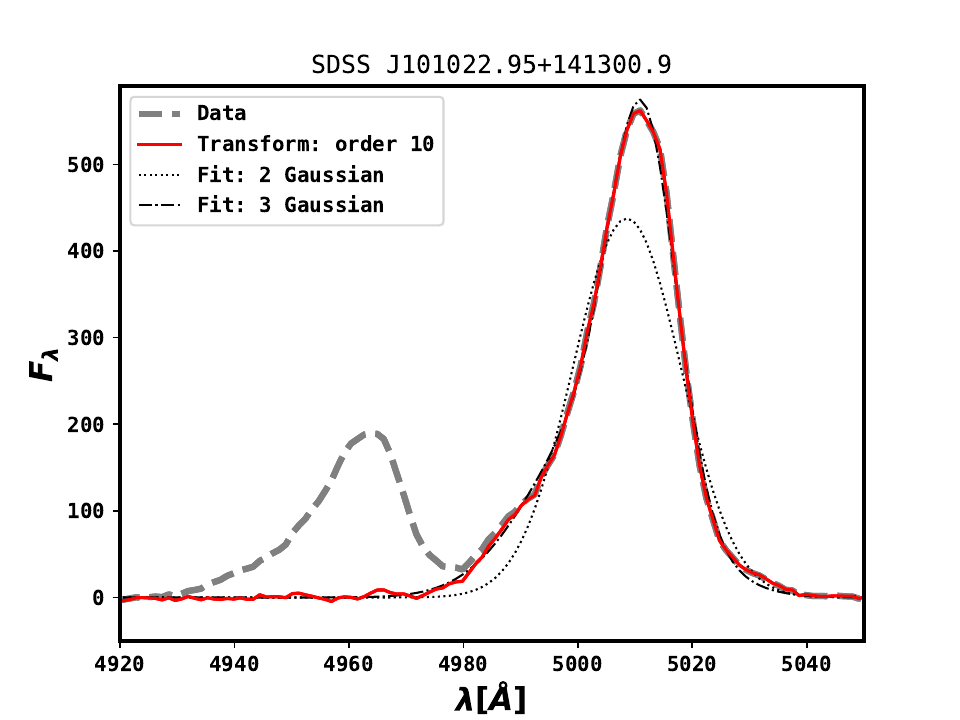}
    \caption{Application of the transform to merged [OIII] emission for a number of SDSS AGN with strongly merged [OIII] emission. The data was 
    background-subtracted before applying the transform. We assumed a line ratio $R=1/3$. For comparison, fits with two and three pairs of Gaussians are shown as dotted and dash dotted black lines respectively. For the fit, we show the solution for the main line, rather than the full doublet for better comparison with the result from the transform; note that fits are performed for the full doublet.}
    \label{fig:example_sdss}
\end{figure*}

Our method is to apply a transformation to $y(x)$ which yields $f_1(x)$. Once $f_1(x)$ has been obtained in  this manner, $f_2(x)$ can then be obtained
trivially by applying Eqn.~\eqref{same_shape_defn}. The shape of each single line can then be analyzed in detail. The transformation is as follows, where $g(x)$ 
is an arbitrary function and $\tilde{g}(x)$ denotes its transform (we use the notation $\tilde{g}$ to denote the transform of a function $g$ throughout this 
work):
\begin{equation}\label{transform_defn}
\tilde{g}(x)\equiv\sum_{m=0}^{\infty}(-1)^mR^mg(x+m\delta).
\end{equation}
Note that the transformation involves superimposing copies of the function $g(x)$, scaled by factors $1$ ,$-R$, $R^2$, 
$-R^3$, $\dotsc$ (which correspond to $m=0$, 1, 2, 3, $\dotsc$), at respective positions $x=0$, $\delta$, $2\delta$, 
$\dotsc$; we emphasise that $R$ and $\delta$ enter into the transformation.
Proof that applying the transformation to $y(x)$ yields $f_1(x)$, i.e. that
\begin{equation}\label{extraction}
\tilde{y}(x)=f_1(x),
\end{equation}
is provided in Appendix~\ref{S:method_derivation}.

With the above in mind we can restate our key result. The transformation, Eqn.~\eqref{transform_defn}, can be 
used to extract the two lines which make up a doublet, assuming the doublet is known to be composed of two similar peaks (Eqn.~\eqref{same_shape_defn}), and the line ratio and wavelength difference are known. Note that the method requires shifting spectra in wavelength, if the sampling of the data does not match the shift applied; interpolation of data points is required.

This is demonstrated in Fig.~\ref{fig:simu_Gaussian} (left panel), where the method is used to resolve a simulated doublet with $R=1/3$, $\delta=1$ and a Gaussian underlying line shape with width $\sigma=0.5$. We use this simulated doublet to test the performance of the method later in Section~\ref{S:considerations}.

We will now apply the method to astronomical data (Section~\ref{S:results}) and then discuss considerations for the use of the method in Section~\ref{S:considerations}.

\section{Example application: Outflows in Type 2 AGN}
\label{S:results}

\begin{figure*}
	\includegraphics[width=\columnwidth]{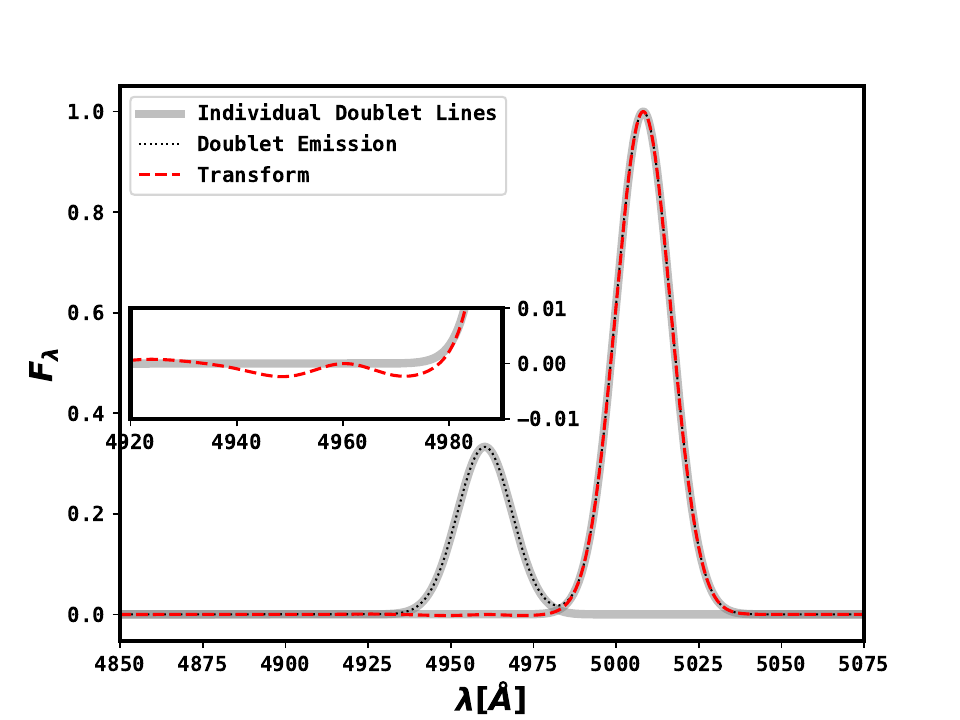}
    \includegraphics[width=\columnwidth]{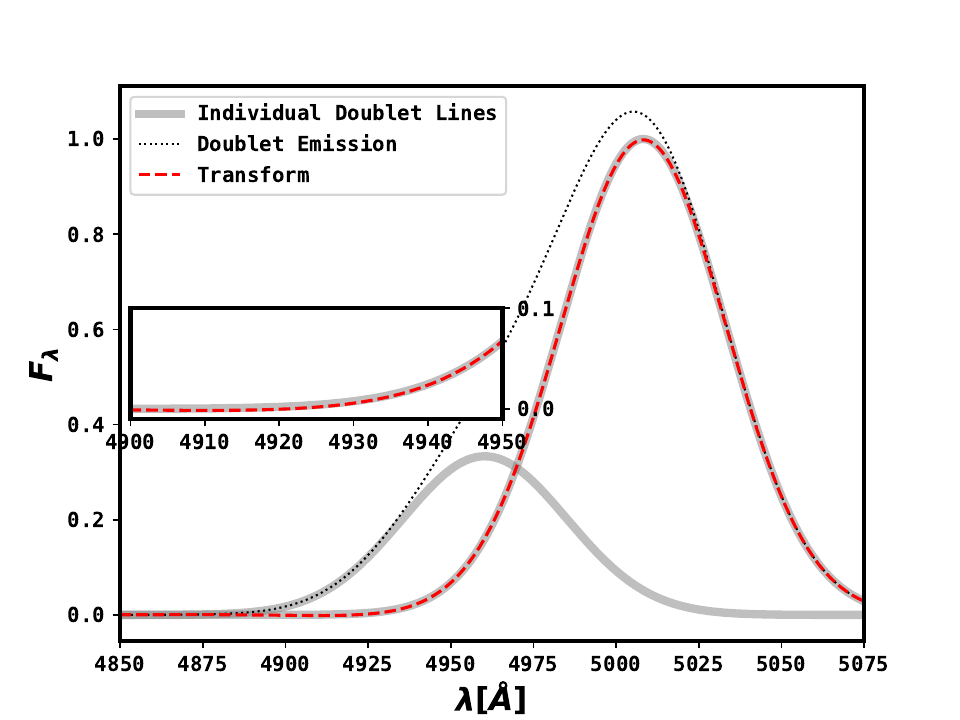}
    \caption{Effect of the fact that the wavelength does not directly onto the velocity. The example simulates typical [OIII] 5007/4959 doublets, the test
    case for this method. The left panel shows a line width of 500~km~s$^{-1}$. Minimal residuals in the tail of the line are apparent. The right panel shows 
    a well merged doublet, where no residuals are seen as a result of the use of wavelength instead of velocity. The line ratio of the simulated doublet is 1/3, the 
    separation between the lines is $\delta=1$ and the width of each line is $\sigma=0.5$.}
    \label{fig:whatisx}
\end{figure*}

To show the strength of the method, we now apply the method to astronomical data. However, as outlined in Section~\ref{S:method}, the method is generally applicable to doublet lines: the method is not just limited to the specific application considered in this section. For instance, here we apply the method to emission lines, although it is also applicable to absorption.

As a test case, we chose AGN with asymmetric [OIII] 5007/4959~{\AA} emission. The lines are separated by $\sim~3000$~km~s$^{-1}$. The [OIII] emission in AGN traces the narrow-line region (NLR). The NLR represents the extended gas ionised by emission from the AGN \citep{antonucci_unified_1993,urry_unified_1995}. The NLR size correlates with the luminosity of the AGN and has a size between $10-10^4$~pc \citep[e.g.][]{schmitt_hubble_2003,liu_observations_2013}. Outflows are common in the NLR and are observed through blueshift with respect to the host galaxy, strong asymmetries in the emission lines \citep[e.g.][]{crenshaw_radial_2009,bae_census_2014}, and detailed kinematic modelling of spatially resolved NLRs \citep[e.g.][]{rupke_integral_2011,crenshaw_feedback_2014,carniani_ionised_2015}. Asymmetries in the observed narrow lines are common since dust in the NLR obscures part of the emission from the outflow, causing line profiles with blueshifted wings \citep[e.g.][]{mullaney_narrow-line_2013,zakamska_discovery_2016}. These outflows form part of a multiphase outflow seen in different gas phases \citep[e.g.][]{cicone_physics_2012,aalto_detection_2012,rupke_integral_2011,rupke_outflows_2005,rupke_multiphase_2013,liu_observations_2013,liu_observations_2013}. The velocity and radial extent of these outflows can be used to estimate outflow rates which can in many cases be well in excess of star formation rates in the host galaxy \citep[e.g.][]{carniani_ionised_2015}. The physics of the large-scale outflows seen in AGN is still under investigation \citep[e.g.][]{fabian_observational_2012,faucher-giguere_physics_2012}, but it is widely believed that such outflows can have a profound effect on the host galaxy by removing gas or terminating star formation in the host galaxy through so-called AGN feedback \citep[e.g.][and references therein]{di_matteo_energy_2005,hopkins_cosmological_2008,fabian_observational_2012}. Studying narrow-line region kinematics, especially at the highest velocities most affected by merged doublet lines \citep{zakamska_quasar_2014,zakamska_discovery_2016}, is therefore of great interest for galaxy evolution studies. NLR emission in AGN is therefore a test case in which line profiles show complex kinematics and cannot be easily approximated by parametric fits, and the wings of the lines tracing the highest velocities are of great interest. The method presented here is therefore ideally suited for this case since it allows decomposition of merged doublets irrespective of the shape of the two lines, even when they are merged (see Fig.~\ref{fig:simu_Gaussian}).

A detailed review of the physics of outflows from both AGN and starburst is beyond the scope of this paper; we refer the reader to the literature for a detailed treatment of the physics of outflows \citep[e.g.][and references therein]{heckman_nature_1990,faucher-giguere_physics_2012,fabian_observational_2012,heckman_brief_2017}.

Here, we apply the method to AGN with a range of narrow line kinematics. Spectra are taken from the Sloan Digital Sky Survey (SDSS) 
\citep{abazajian_seventh_2009} covering the rest frame wavelength range 3800--9200~{\AA}\ and have a spectral resolution of $R_{\text{spec}}\approx$1850--2200. We selected Type 2 AGN to avoid contamination from iron emission \citep[e.g.][]{kovacevic_analysis_2010,vestergaard_empirical_2001} or broad H~$\beta$ emission, since a contaminating line is a complication which the method was not designed to account for. We discuss the effect of a contaminating line on the results of the method in more detail in Section~\ref{S:linecontamination}.

We applied the method to AGN for which [OIII] is in the SDSS spectral window (Z $\leq$ 0.7) and a significant detection of [OIII] is present, leaving $>1000$ spectra, depending on the cut-off used for [OIII] detection. Of those, a large fraction were visually inspected to check for obvious residuals; the transform was found to be robust. Here we present, for the sake of brevity, results for only four of the AGN we considered, chosen to exhibit a range of line profiles, thus illustrating that the method works irrespective of the underlying line shape. Specifically, we choose three sources showing the strongly skewed emission lines to illustrate the performance of the transform for merged emission lines. The chosen objects show a range of S/N, illustrating the performance of the transform under different levels of noise. Additionally, we show a double-peaked [OIII] emitter to demonstrate that the method recovers even complex line profiles without residuals.

\begin{figure*}
    \includegraphics[width=\columnwidth]{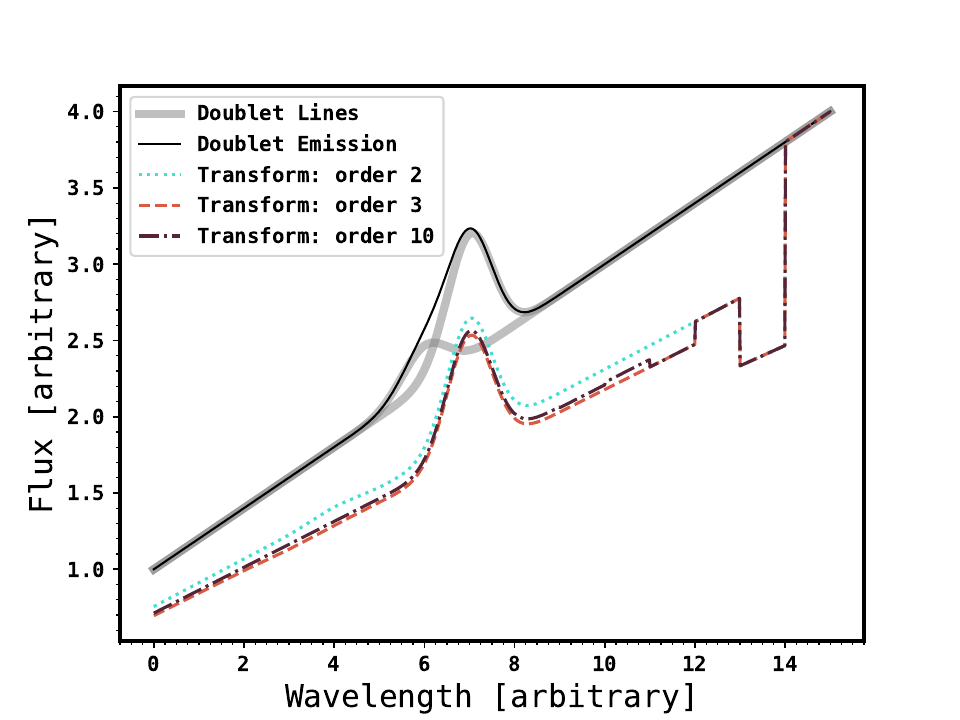}
    \includegraphics[width=\columnwidth]{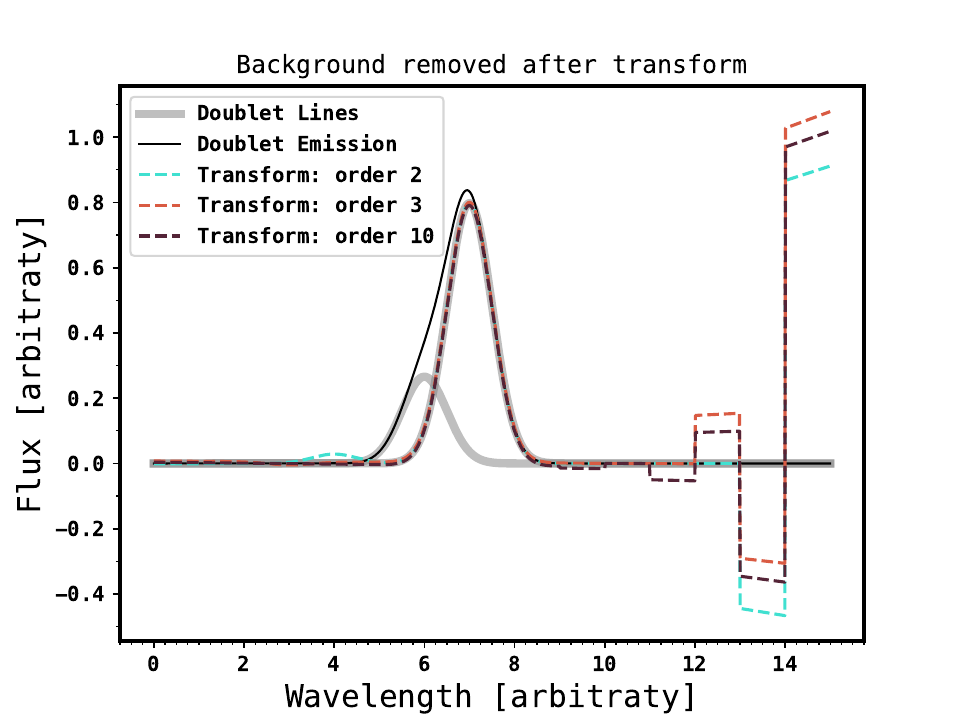}
    \caption{Simulations showing the performance of the transform for imperfect background subtraction. Left: A linear background has been added to the signal 
    (the model doublet, described in the main text).  
     Right: Data to which the transform has been applied, identical to the data in the left, but with the background is fit and subtracted. The extracted line
     shapes are correct, regardless of the presence of a linear background in the signal. The line ratio of the simulated doublet is 1/3, the separation between 
     the lines is $\delta=1$ and the width of each line is $\sigma=0.5$.}
    \label{fig:simu_bg}
\end{figure*}

Before applying the method, continuum emission was removed by locally fitting a linear function, and then subtracting the continuum. Having a (flat) non-zero continuum contribution does not adversely affect recovery of the emission line shapes (see Section~\ref{S:background}, Fig.~\ref{fig:simu_bg}); the transform was applied to non-continuum subtracted data as well and shown to be reliable. We show the continuum subtracted data to allow better assessment of the residuals.

The aforementioned four [OIII] doublets, as well as the results of applying the transform to them, are shown in Fig.~\ref{fig:example_sdss}.
The [OIII] doublets are significantly merged in some of the cases shown here. 
The transform separates the two lines, revealing the underlying line shape. Note that the transform does not alter the peak of the line. The transform allows recovery of the tail of the line, revealing the high-velocity structure in the outflow. The example of the double-peaked emission line (bottom left panel) shows that the transform can recover even complex line shapes reliably. The method therefore enables study of the complex kinematics in powerful outflows without having to rely on fitting models with a large number of free parameters.

For comparison, the results of applying a conventional fitting procedure to resolve the doublets are also shown in Fig.~\ref{fig:example_sdss}. The fitting procedure we employed is similar to those commonly employed in the field, and involved modelling each line of the doublet as either two \citep[similarly to e.g.][]{mullaney_narrow-line_2013} or three Gaussians, with the line ratio and wavelength difference of the lines coupled and fixed. The standard \texttt{scipy}
function \texttt{curve\_fit} was used for minimisation.

The top row in Fig.~\ref{fig:example_sdss} corresponds to the most strongly-merged doublets. It is not obvious what the `true' underlying line shape is for these cases. Hence we first discuss weakly-merged doublets in the figure (bottom row), where the underlying line shape is more `obvious'. Here we see that the line shapes obtained from the transform appear correct: they are in excellent agreement with those of the 5007~{\AA} line. By contrast, the two-Gaussian-component fits fail to reproduce key features of the line shape. While the three-component fits reproduce the line shape better, significant convergence issues start to emerge; all four doublets required the starting parameters to be modified to achieve converged fits. With regards to the strongly-merged doublets (top row), similarly to the weakly-merged doublets, the transform yields a line shape which closely resembles that of the 5007~{\AA} line, while obtaining reasonable-looking fits with the fitting procedure is more difficult. Additionally, for the noisier line profile (upper left), the three-component fitting procedure required considerable modification of starting parameters to achieve a good fit. By contrast the transform performs equally well for lower S/N data. This can be difficult to achieve in large samples.

This comparison highlights the key strength of the transform over conventional fitting procedures: it reproduces line shapes without the need for carefully choosing input parameters or dealing with issues such as convergence or choice of initial conditions which bedevil conventional fitting procedures. The transform also remains robust for lower S/N data that can be challenging for modeling. Moreover the transform involves only a single operation on the data array per order, regardless of the complexity of the line shape. Hence the transform will outperform fitting procedures which involve optimisation over many free parameters -- we note that using many free parameters are necessary to capture complex line shapes.

All further applications of the transform in this work are to simulated data, which has the advantage over real data that the correct solution is known. Simulated data thus affords a more rigorous test of the transform. Specifically, we test the performance of the transform in the presence of continuum contamination, line contamination, noise and when the line ratio has an error.

\section{Considerations for use of method}\label{S:considerations}

The transformation defined in Eqn.~\eqref{transform_defn} can be applied to a doublet $y(x)$, comprised of two merged, geometrically similar lines, in order to extract the individual lines $f_1(x)$ and $f_2(x)$. We showed in Fig.~\ref{fig:simu_Gaussian} that this method recovers ideal, noise-free data. In this section, we will discuss the numerical implementation of the method, as well as the effect of issues in real data: specifically, contamination from noise, background, and errors in the line ratio $R$ and peak separation $\delta$.

It should be noted that the spectra this method will be applied to are given either as a function of wavelength or frequency. Earlier, we explicitly assumed that the lines were geometrically similar in wavelength. The emission lines are, however, similar in velocity $v$, not wavelength $\lambda$. Now, for a source with rest wavelength $\lambda_{\text{rest}}$, the mapping between observed wavelength $\lambda$ and source velocity $v$ is given by (for $v\ll c$)
\begin{equation}
\lambda = \left(\frac{v}{c} + 1 \right)\lambda_{\text{rest}} 
\end{equation}
where $c$ is the speed of light. Hence, since the two lines have different values of $\lambda_{\text{rest}}$, there is no mapping between $\lambda$ and $v$ common to both lines. A result of this is that if the two lines are geometrically similar in velocity then they are not geometrically similar in wavelength, and vice versa. Hence our earlier assumption of similarity in wavelength, which underpins the transform, is invalid. However, the transform remains valid for all intents and purposes if $\lambda_1 \approx \lambda_2$. This is shown in Fig~\ref{fig:whatisx}, where doublets, geometrically similar in velocity, have been transformed into the wavelength frame before the transform is applied to extract the underlying line shape.  It can be seen from the figure that the resulting residuals are minimal if the velocity separation of the line is comparable to the width of the line (right panel), and only become noticeable for line widths much smaller than the separation (left panel). Even in this case, the errors are negligible ($\sim 1\%$ for a realistic case shown in Fig.~\ref{fig:whatisx}) and affect only the tail of the lines. These results imply that if the method is to be used for doublets comprised of well-separated lines, each line should separately be transformed to the velocity frame. However, the case of very well separated lines is not the intended use case of this transform, and discussion of this is beyond the scope of this paper.

We now discuss how limitations such as noise, background, and errors in the assumed line ratio, affect the performance of the transform.

\subsection{Truncation and convergence}
\label{S:trunc_conv}

As can be seen from its definition, Eqn.~\eqref{transform_defn}, the transformation involves an infinite number of iterations. Each term $m$ in the sum corresponds to a shifted and scaled version of $g(x)$, with $R^m$ the scale factor and $m\delta$ the shift applied to $g(x)$. In numerical implementations of the transformation, it is of course necessary to truncate the infinite summation. We refer to the term $m=M$ at which the summation is truncated as the \emph{order}, i.e. terms in Eqn.~\eqref{transform_defn} with $m>M$ are omitted. Thus in numerical implementations Eqn.~\eqref{transform_defn} becomes
\begin{equation}\label{transform_order}
\tilde{g}(x)\equiv\sum_{m=0}^{M}(-1)^mR^mg(x+m\delta).
\end{equation}
We henceforth assume that the transform is defined by the above equation as opposed to Eqn.~\eqref{transform_defn}.

For application to the data, we therefore need to know by which $M$ the calculation should be truncated so that convergence is achieved. Recall that convergence and the order of the transformation have been demonstrated in Fig.~\ref{fig:simu_Gaussian} (left panel). This doublet was designed to be representative of real-world [OIII] doublets (see Section~\ref{S:results}). We use the same doublet when we examine the effects of background, noise, and uncertainty in $R$ and 
$\delta$ below.

Fig.~\ref{fig:simu_Gaussian} shows the working of the method, as described in Section~\ref{S:method}: the data is shifted and subtracted from itself, the increasing iterations correcting for the over-subtraction (which gives rise to the residuals shown in the inset figure). As can be seen from the figure, the transform corrects for residuals of the previous iterations further and further from the peak of the main line. Choosing a particular order $M$ means that the transform is converged over a wavelength range $\approx M\delta$ downstream from the main line (where downstream means in direction of the minor line).  The order should therefore be chosen keeping in mind the wavelength range of interest. Similarly, to avoid edge effects, if applying the transform to the order of $M$, a background of width $M \delta$ should be included (this is discussed in detail in Section~\ref{S:background}).

Note that the transform converges only for $R<1$. The main line should therefore always be chosen so that the line ratio is smaller than one, which can be achieved by reordering the lines. $R=1$ constitutes a special case. Here, the transform does not converge since, unlike for the case of $R<1$, successive  iterations always perform corrections of the same magnitude. However, the transform still yields converged results over a wavelength range $M\delta$ from the peak of the main line.  Choosing a suitable $M$, as described generally above, is therefore of greater importance for $R=1$.

In summary, the order $M$ should be chosen keeping in mind the wavelength range $\Delta \lambda$ of interest so that $\Delta\lambda \approx M \delta$ ($M \delta$ both up- and down-stream of the main line).

We emphasise that due to the simplicity of the method, computation time is minimal, irrespective of order. Furthermore, note that while we have modelled the line shape here as a Gaussian, we emphasise that our method is general in that it works for \emph{any} line shape; the method makes no assumption about the underlying line shape, e.g. that it is a Gaussian. 


\subsection{Continuum contamination}
\label{S:background}

Generally, spectral data will have a non-zero continuum. This can be fit for and subtracted, but residuals are likely to remain. Here, we discuss how continuum contamination affects the performance of the transform. In the following we refer to continuum contamination from any source as background.

To account for noise or a background, Eqn.~\eqref{I_defn} can be generalised to
\begin{equation}\label{I_better_defn}
y(x)=f_1(x)+f_2(x)+\eta(x),
\end{equation}
where $\eta(x)$ describes the contribution to the observed doublet due to noise and a background. We now investigate how the addition of $\eta(x)$ affects the ability of the transform to extract the singlet $f_1(x)$ (Eqn.~\eqref{extraction}).

We begin with a general result. Applying the transform to Eqn.~\eqref{I_better_defn}, and exploiting Eqn.~\eqref{extraction} and the linearity of the transform (see Appendix~\ref{Appendix:linearity}), it follows that 
\begin{equation}\label{extraction_gen}
\tilde{y}(x)=f_1(x)+\tilde{\eta}(x).
\end{equation}
Hence if the doublet exhibits noise or a background, then applying the transform still obtains the singlet $f_1(x)$ but with the transform of the background, $\tilde{\eta}(x)$, added to it.

Ideally the background would be perfectly subtracted before the transform is applied to the doublet. If it is not, we can generally approximate the background as linear across the doublet. For this reason we now consider how the presence of a linear background affects the results. A linear background corresponds to $\eta(x)=Ax+C$, where $A$ and $C$ are constants, and in Appendix~\ref{Appendix:background} it is shown that its transform is given by
\begin{equation}\label{background_lin}
\tilde{\eta}(x)=A'x+C',
\end{equation}
where
\begin{equation}
A'=A\frac{1-(-R)^{M+1}}{1+R}
\end{equation}
and
\begin{equation}
C'=C\frac{1-(-R)^{M+1}}{1+R}- A\delta\frac{R}{(1+R)^2}\Bigl[1+M(-R)^{M+1}-(M+1)(-R)^M \Bigr].
\end{equation}
(Recall that $M$ is the order of the transform; see Eqn.~\eqref{transform_order}). Hence a linear background becomes a linear background in the transform, albeit shifted and rescaled such that the gradient $A'$ and intercept $C'$ differ from the background in the raw data.

Crucially, the nature of the line obtained from the transform is unaffected by the presence of the linear background; the shifted and rescaled background is simply added to the singlet $f_1(x)$ extracted from the transform. Thus $f_1(x)$ (and hence $f_2(x)$ via Eqn.~\eqref{same_shape_defn}) can be extracted from $\tilde{y}(x)$ by subtracting the transformed background (see Eqn.~\eqref{extraction_gen}).

Fig.~\ref{fig:simu_bg} shows the effect of applying the transform to our model doublet in the presence of a linear background. As can be seen from the left panel of Fig.~\ref{fig:simu_bg}, applying the transform to the data yields the single line plus a linear background with a different gradient and intercept, as predicted by the above equations. Note, however, that there are edge effects due to the fact that the  simulated array is not of infinite length; for increasing orders of the transform, we observe edge effects due to the fact that the shifted data contains no information beyond the limit of the data, and this boundary is shifted further and further left. This effect is visible only when a non-negligible background is present. This should be taken into account when choosing the wavelength range as well as the order $M$ of the transform (see also Section~\ref{S:trunc_conv}). Specifically, if applying the transform with order $M$, a background of width $M \delta$ should be included upstream to avoid the edge effect adversely affecting features of interest.

In the right panel of the figure the linear background in the transformed data has been fit and removed, leaving a single line which is indistinguishable from the `true' line. Imperfect background subtraction therefore does not affect the transform. The transform is robust to background contamination.

\subsection{Noise}
\label{S:noise}

Noise will be present in all astronomical datasets. We now consider the effect of noise. Specifically, we consider \emph{Gaussian additive noise}: $\eta(x)$ is a  random variable with mean 0 and variance $\sigma^2$, i.e. $\langle\eta(x)\rangle=0$ and $\langle\eta(x)^2\rangle=\sigma^2$ for all $x$. In this case it can be shown (see Appendix~\ref{Appendix:noise}) that, for all $x$, the mean and variance in $\tilde{\eta}(x)$ are given by
\begin{equation}\label{noise_mean}
\langle\tilde{\eta}(x)\rangle = 0
\end{equation}
and
\begin{equation}\label{noise_var}
\langle\tilde{\eta}(x)^2\rangle = \sigma^2\Biggl[\frac{1-R^{2(M+1)}}{1-R^2}\Biggr]
\end{equation}
respectively, where we have assumed that $R<1$. Eqns.~\eqref{noise_mean} and \eqref{extraction_gen} imply that Gaussian additive noise does not `distort' the singlet obtained from the transform:
\begin{equation}
\langle\tilde{y}(x)\rangle=f_1(x).
\end{equation}
However Eqn.~\eqref{noise_var} implies that the fluctuations in the noise are always magnified by the transformation, with larger magnifications as $M$ is increased (with a limiting value of $1/(1-R^2)$ as $M\to\infty$) or as $R$ approaches 1. Though, as mentioned above, this does not affect the shape of the single line obtained from the transform, and thus Gaussian additive noise will not cause biases in line shape analysis. 
To restate, the transformation will recover the correct line shape in the presence of Gaussian additive noise (and assuming no other complications such as contamination from other emission lines, e.g. contamination from other emission lines) regardless of the signal-to-noise ratio.

For $R=1$ Eqn. \eqref{noise_mean} still applies, but $\langle\tilde{\eta}(x)^2\rangle$ is instead given by (see Appendix~\ref{Appendix:noise})
\begin{equation}\label{noise_var_R_1}
\langle\tilde{\eta}(x)^2\rangle = \sigma^2(M+1).
\end{equation}
This diverges as $M\to\infty$. However, this problem is sidestepped since, as discussed in Section~\ref{S:trunc_conv}, in practice we consider only a finite wavelength range, and hence it is sufficient to use a finite $M$. 

However, while Eqns.~\eqref{noise_var} and \eqref{noise_var_R_1} imply that the magnitude of the the noise in the transform is the same for all $x$, these equations have been derived assuming the presence of noise at all wavelengths. In practice we would be considering a finite wavelength range, and the obvious course of action is to assume that there is no noise outwith the considered range. In this case noise is not present at all wavelengths, and Eqns.~\eqref{noise_var} and \eqref{noise_var_R_1} break down. The result is that $\langle\tilde{\eta}(x)^2\rangle$ becomes $x$-independent. This is an edge effect which could significantly complicate the task of correctly fitting the underlying line shape. This problem can, however, be sidestepped by using a sufficiently large wavelength range so that the line shape is converged before the distance $M\delta$ is comparable to the distance from the main peak to the edge.

In Fig.~\ref{fig:simu_Gaussian} (right panel) the transform is applied to our model doublet, but with noise with standard deviation $\sigma_{\text{noise}}=0.01$ added to the data.
Crucially, the underlying line shape is still recovered reliably despite the noise. Moreover, as in the noise-free data, $M=10$ is sufficient to obtain converged results. As mentioned above, the magnitude of the noise will be magnified by the transform. For the $R$, $\delta$, $M$ and $\sigma_{\text{noise}}$ in the simulated data, Eqn. \eqref{noise_var} predicts that the noise in the transformed data will have standard deviation $\tilde{\sigma}_{\text{noise}}=0.0106$, which corresponds to a 6\% magnification. This increase is too small to be noticeable in the figure. Further results for the case $R=1$, where the magnification in the noise is substantial, are presented in Appendix~\ref{Appendix:R1}.

\begin{figure*}
	\includegraphics[width=\columnwidth]{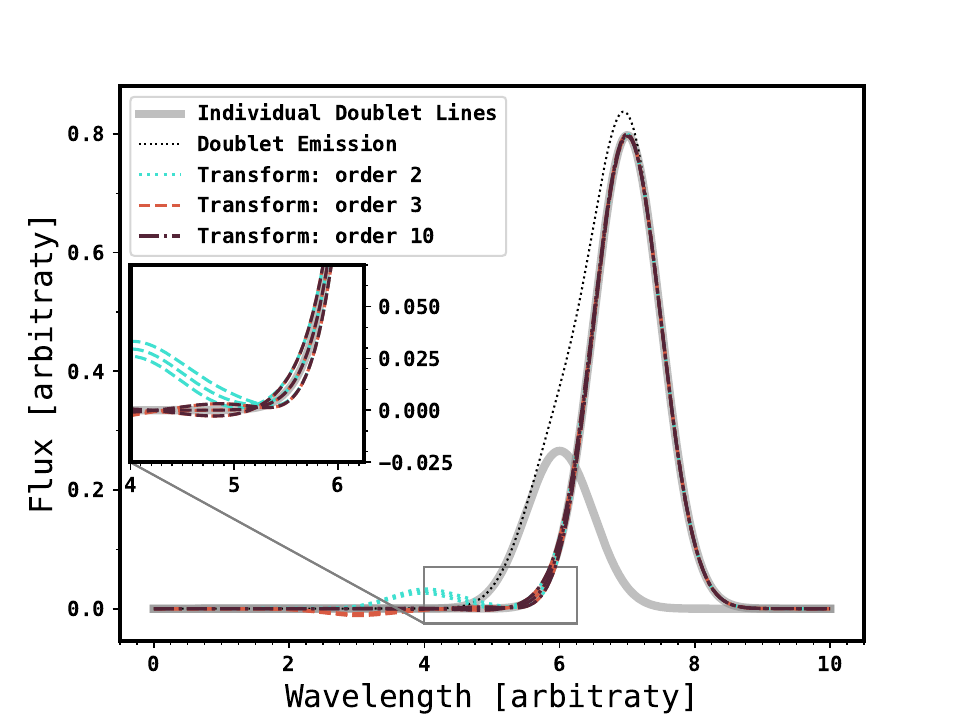}
    \includegraphics[width=\columnwidth]{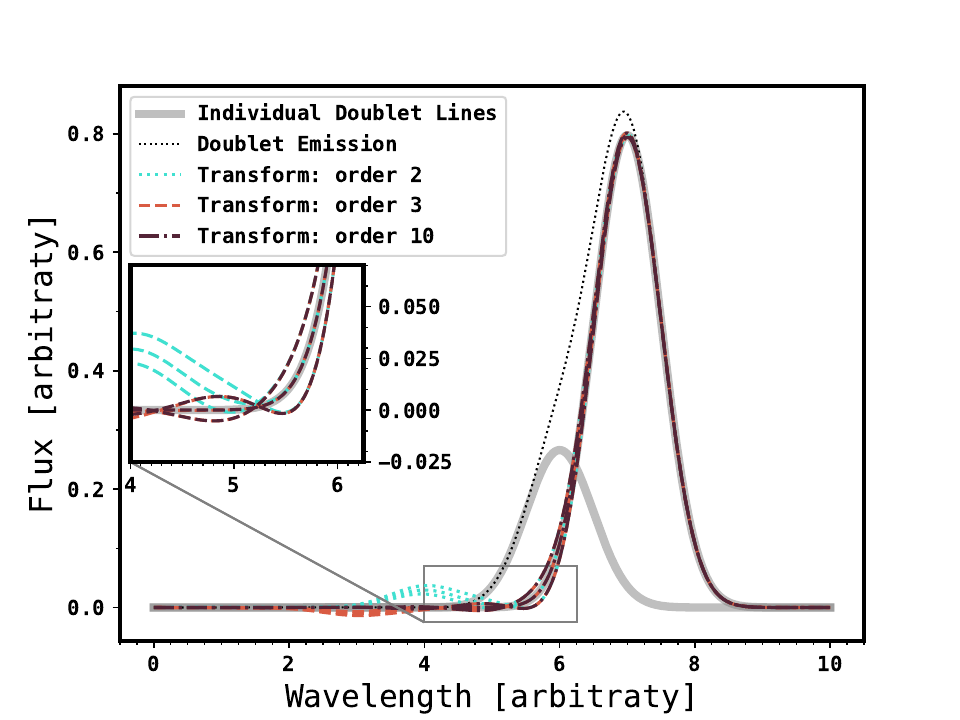}
    \caption{Effect of applying the transform with an incorrect line ratio $R$. The left panel corresponds to a 5\% error in $R$, the right one a 10\% error. In 
    both cases the `true' line ratio of the simulated doublet is $1/3$, the separation between the lines is $\delta=1$, and the width of each line is 
    $\sigma=0.5$.}
    \label{fig:R_uncertainty}
\end{figure*}

\subsection{Uncertainty in line ratio and peak separation}
\label{S:delta_R_uncertainty}

The appropriate values of $R$ and $\delta$ to use in the transform (Eqn.~\eqref{transform_order}) might not be known exactly. Hence it is desirable to know the effect of applying the transform with slightly incorrect values of $R$ and $\delta$. In practice the line separation will be known to a high accuracy; it is the line ratio which has the dominant source of uncertainty. Hence we focus on the effect of uncertainty in $R$ here.

Fig.~\ref{fig:R_uncertainty} shows the effect of applying the transform to our model doublet, but using slightly `wrong' values of $R$, 
namely 5\% and 10\% above and below the true value of $R$. It is clear from the figure that the effect of using slightly incorrect values 
of $R$ here is small. To elaborate, here, when the order of the transform is sufficiently high to yield convergence (i.e. by order 
$M=10$), using a value of $R$ which differs from the true value by 10\% yields errors with magnitude $\lesssim 2$\% of the maximum in the 
doublets. However, such small errors may not result for all doublets. To facilitate further investigation of this, equations 
quantifying the effect of using incorrect $R$ and $\delta$ in the transform are provided in Appendix~\ref{Appendix:delta_R_uncertainty}.

\subsection{Contamination from emission lines}
\label{S:linecontamination}

Doublet emission might be contaminated by unassociated line emission, either from the astrophysical source itself or from sky emission. The resulting transform will consist of the transform of the doublet, with the the transform applied to any contaminating line emission superimposed on top of it. (This follows from the linearity of the transform; see Appendix \ref{Appendix:linearity}). In other words, the contaminating line will be shifted and subtracted from the dataset, resulting in residuals in the transform with separation $\delta$ and amplitude $R$ times the original amplitude of the contaminating line.

\begin{figure*}
    \includegraphics[width=\columnwidth]{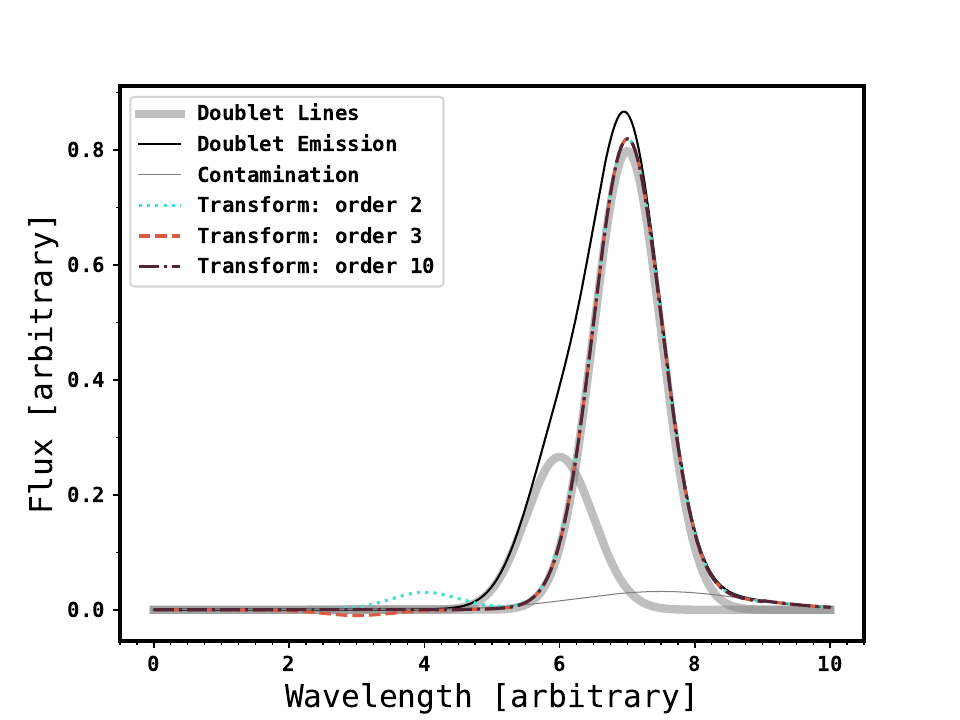}
    \includegraphics[width=\columnwidth]{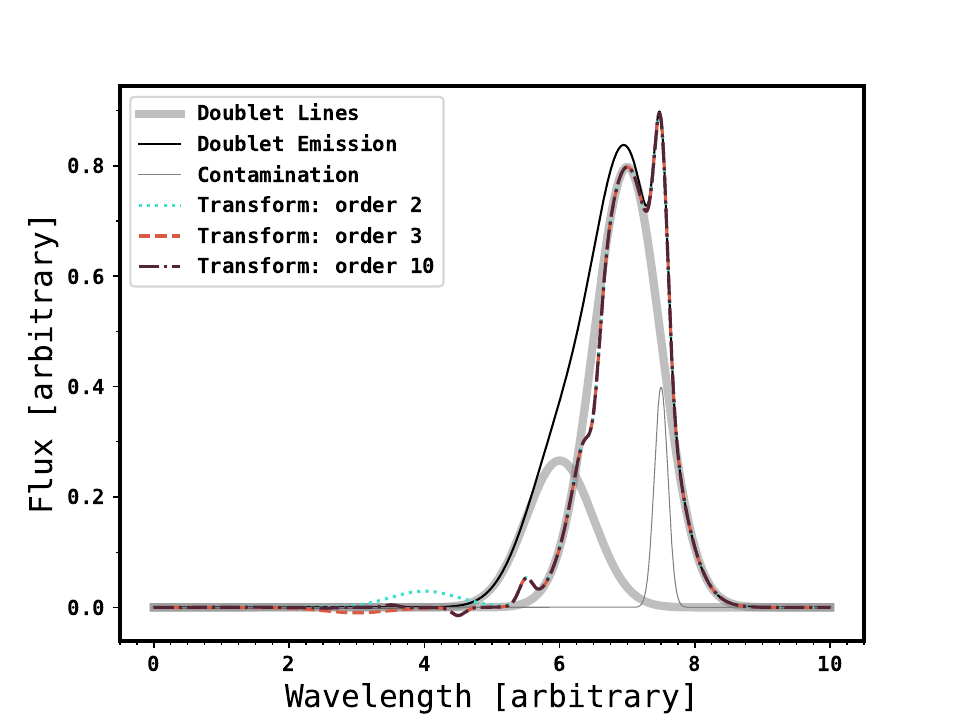}
    \caption{Simulations showing the effect of line contamination on the transform. As in Fig~\ref{fig:simu_Gaussian}, $R=1/3$, $\sigma=0.5$ and $\delta=1$. We show two extreme cases, the left panel shows contamination from a line much broader than the doublet line ($\sigma=1.25$) and the right panel shows contamination from a narrow emission line ($\sigma=0.1$). Both contaminating lines have 10\% of the flux in the main line. We see that the broader emission line does not strongly affect the transform, with the line shape recovered well. The narrow emission line results in residuals with decreasing amplitude with separations $\delta$, as the transform shifts and subtracts the contaminating line from the contaminated doublet.}
    \label{fig:linecontamination}
\end{figure*}

This is illustrated in Fig.~\ref{fig:linecontamination}, for cases of broad (left panel) and narrow (right panel) contaminating lines with 10\% flux of the main line. It can be seen that the residuals resulting from the broad contaminating line are minimal; while the transform does not remove the contaminating line, it does not increase the error resulting from contamination further. By contrast the narrow contaminating line introduces noticeable residuals at intervals of $\delta$ from the location of the contaminating line, which decrease in amplitude with distance from it. As for the broad line, the transform does not remove the narrow contaminating lines. The transform will therefore result in a residual in the presence of line contamination, but the amplitude of this is always smaller or equal to the initial line.

All of the above observations should be borne in mind if the transform is to be applied in the presence of contaminated line emission.

\section{Summary and Discussion}
\label{S:discussion}

Here, we have presented a method (Section~\ref{S:method}) that allows the non-parametric separation of merged doublet emission lines when the line ratio is known. The method is generally robust for noisy data and recovers the emission line with high precision without introducing significant noise (Section~\ref{S:considerations}). The transform is also robust to background contamination.

More specifically, we have presented mathematical derivations and simulation results which show that the method is robust in the presence of a linear background (Section~\ref{S:background}), and does not inflate noise considerably (Section~\ref{S:noise}) unless the line ratio is close to 1. The method is also numerically fast, enabling it to be easily applied to large datasets for further processing without the need to fit emission lines. Another appealing feature of the method is that it is easily implemented; \textsc{python} source code is included in Appendix~\ref{python}. We have also applied the method to extreme AGN-driven outflows (see Section~\ref{S:results}, Fig.~\ref{fig:example_sdss}) and show that it allows line shapes to be recovered, including the scientifically interesting high-velocity wings that trace the most high-velocity structures in the outflow.

To conclude this paper we summarise the requirements for reliable use of of the method, and suggest possible future applications.

\subsection{Requirements for use of method}
\label{S:reqs}

The method presented here can be a powerful tool for decomposing doublet emission. However the following limitations should be kept in mind when 
applying the method.
\begin{itemize}
\item Sampling: The data needs to be sampled appropriately both for any features to be recovered as well as the wavelength difference $\delta$. 
\item Noise levels and S/N: For low values of $R$, noise is only mildly increased by the transform. The noise levels in the transform diverge as $R$ approaches 1 for infinite number of iterations in the method $M$. However, 
since the wavelength range $\Delta\lambda$ used in practice is limited, and therefore the number of iterations required for convergence of the transform is limited to $M\approx \Delta \lambda / \delta$, the noise level does not reach infinity. Noise levels should be considered for low signal-to-noise data and line ratios $R\sim 1$. The increase in 
noise level is discussed in Section~\ref{S:noise} and in more general terms in Appendix~\ref{Appendix:noise}. There is therefore no formal signal-to-noise requirement for the use of the method, and while the transform will increase noise levels for typical cases ($R<1$, $M<10$), the increase is small, only by a few per cent.
\item Continuum background subtraction: Flat as well as linear backgrounds do not affect the transform; see Fig.~\ref{fig:simu_bg} as well as Section~\ref{S:background}
and Appendix~\ref{Appendix:background}.
\item Line ratios: The line ratio needs to be known; if the assumed line ratio is incorrect, the transform does not return the underlying line shape. The error
introduced is discussed in Section~\ref{S:delta_R_uncertainty} and in Appendix~\ref{Appendix:delta_R_uncertainty}. For small errors $\sim 5\%$ in the line ratio, the resulting errors in the transform are small ($<1\% $; see Fig. \ref{fig:R_uncertainty}).
\item Special case -- line ratio $R=1$: In the case of line ratio 1, the transform does not converge on an infinitely large wavelength range. However, the transform still converges over a limited wavelength range. In this case $M$ should be chosen carefully; see Section~\ref{S:trunc_conv} as well as \ref{S:noise}.
\item Contamination by other emission lines (e.g. FeII or H~$\beta$ in case of the [OIII] doublet) cannot be accounted for using the method. However, this limitation equally applies to any other method that can be used to decompose merged doublets. The resulting residuals are outlined in Section~\ref{S:linecontamination}.
\end{itemize}

Under these conditions, the method allows decomposition of a doublet for further non-parametric characterisation, such as percentiles or skew measurements. The method also allows separation of stacked spectra, which might not follow ideal parametric functions. It allows the direct comparison of different doublet lines, such as [OIII] and [OII], without requiring a parametrization. The method is also ideally suited for visualisation purposes for strongly-merged doublets, as shown in Section~\ref{S:results}, Fig.~\ref{fig:example_sdss}. The strength of the method is its speed and simplicity, which makes it ideal for application to large samples.

\subsection{Future applications}
\label{S:use}

As explained above, the method can be used for any doublet emission with suitable data. Here we list a few scientific cases for which the data would be suitable.
\begin{itemize}
\item \,[OII] 3727/3730~{\AA}: This doublet is heavily merged, with a separation of only $\sim 250$~km~s$^{-1}$. [OII] however is used for kinematic studies at redshifts for which [OII] is shifted outside of the optical range \citep{patricio_kinematics_2018}. The line ratio is insensitive to density above a density threshold of $10^4$~cm$^{-3}$ and so can be assumed to be fixed for most cases \citep{draine_physics_2011}.
\item MgII 2796/2803~{\AA} is seen in absorption in outflows from AGN. With a velocity separation of only $\sim 750$~km~s$^{-1}$, merged emission lines can be common. If the emission can be assumed to be either optically thick or thin, the line ratio is known \citep{kovacevic_analysis_2010}.
\item Velocity delay maps are commonly used to study the kinematics of the broad line region (BLR) of AGN \citep{welsh_echo_1991}. BLR lines show complex profiles; for MgII, where the doublet has a velocity separation of $\sim 750$~km~s$^{-1}$, the doublet causes significant smearing. If the line is optically thick, the method presented here can be used to decompose BLR emission since the line ratio is known. 
\item Absorption line kinematics: while we have discussed and visualised the method in terms of emission lines, the method is equally suitable for absorption-line doublets (such as MgII or NaD outflows). As for the case of emission lines, contamination from other lines will affect the transform and result in residuals, see Section~\ref{S:linecontamination} for details.
\item The transform is also ideal for quick visualisation of merged doublet emission without requiring fitting.
\end{itemize}
Naturally, this list is not complete. Moreover, since the problem of separating doublets comprised of identically-shaped singlets is encountered in many fields of science, there is the prospect of the method being applied generally outside of astrophysics.

In summary, we have presented a simple non-parametric method for resolving merged doublet emission. The method is robust to linear background contamination. The method can be used for a wide range of science cases as well as visualisation. \textsc{python} code for implementation is given in Appendix~\ref{python}.

\section*{Acknowledgements}

We thank the referee for constructive comments that have improved the quality and clarity of the manuscript. We thank Keith Horne, Mike Goad and Kirk Korista for useful comments and discussion. Contributions to the paper: CV performed testing and identified the astrophysical test and use cases. TLU had the idea for and derived the proof of the method. CV\& TLU co-wrote the paper. NM and MT were involved in testing of the method. 

\bibliographystyle{mnras}
\bibliography{Doublets}


\appendix

\section{Python implementation of algorithm}
\label{python}

The function to implement the transform in Python is presented below. As an input, the function takes the wavelength array, the flux array, $\delta$, $R$, and optionally the order $M$ (denoted in the code below as \texttt{delta}, \texttt{R} and \texttt{order}, respectively). \texttt{flux.scipy.interpolate} is used to interpolate the spectrum and allow evaluation of the spectrum after the wavelength shift is applied.
However such interpolation is not an essential aspect of the method, and it would be possible to write a function which did not rely on this.

\begin{lstlisting}[language=Python, linewidth=\columnwidth]
import numpy
import scipy.interpolate
 
def twin_peaks(wl, flux, delta, R, order=10):
    """
    A function to separate doublet lines.
    The outflows are not what they seem.
    x: input wavelength array
    y: input flux array
    delta: offset between the two lines,
    same unit as x (positive if shorter
    wavelength line has lower flux)
    R: ratio (<1) between emission lines,
    order: order to which to correct for,
    default is 10, note that higher order
    have no effect if wavelength range is
    shorter than order*offset
    returns: flux after transform
    """
    #Interpolation to allow evaluation of the
    #spectrum at different wavelength after
    #shifting the array
    rawf = scipy.interpolate.interp1d(wl, flux,
    bounds_error=False, fill_value=0)
    outf = numpy.zeros_like(flux)
    outf += flux
    for i in range(1, order+1):
    	#Spectrum is shifted and added
        #to original dataset.
    	outf += (-1)**i * R**(i)
     		* rawf(wl + i * delta)
    return outf
\end{lstlisting}

We now outline how this function can be applied.
As a first step, the spectrum needs to be loaded. The function requires a wavelength and flux array with sufficient wavelength resolution. No error data are required. The wavelength and flux arrays can be assigned in the usual manner, e.g.
\begin{lstlisting}[language=Python, linewidth=\columnwidth]
wl, flux = ...
\end{lstlisting}
(where \texttt{...} signifies code specific to the problem at hand).
In this example, we are applying the transform to an [OIII] doublet with a wavelength difference of 47.945~{\AA} and a line ratio of 1/3, as in the test cases shown in Section \ref{S:results}.
Since the stronger line has a longer wavelength, the wavelength difference \texttt{delta} must be given as \textit{positive}. For a doublet were the shorter wavelength line is stronger, the wavelength difference \texttt{delta} would be given as \textit{negative}. The line ratio \texttt{R} should always be $<1$. Accordingly, for this case \texttt{delta=47.945} and \texttt{R=1./3} in the function.
The only other information required by the transform is the order. In this example, we choose 10, which will apply the transform over 10*\texttt{delta}$\approx$500~{\AA}. Hence transform in this example is applied as follows:
\begin{lstlisting}[language=Python, linewidth=\columnwidth]
transform = twin_peaks(wl=wl, flux=flux,
	delta=47.945,
	R=1./3,
    order=10)
\end{lstlisting}
The function returns the transformed spectra, \texttt{transform}, with the same array shape as the input spectrum. A comparison between the original spectrum and transform could be plotted as follows:
\begin{lstlisting}[language=Python, linewidth=\columnwidth]
import pylab
pylab.plot(wl, flux)
pylab.plot(wl, transform)
\end{lstlisting}

\section{Further results}\label{S:more_theory}

\subsection{Derivation of Eqn. \eqref{extraction}}
\label{S:method_derivation}
To derive Eqn.~\eqref{extraction}, we first substitute Eqn.
\eqref{same_shape_defn} into Eqn.~\eqref{I_defn}. This gives
\begin{equation}
y(x)=f_1(x)+Rf_1(x+\delta),
\end{equation}
which becomes
\begin{equation}
f_1(x)=y(x)-Rf_1(x+\delta)
\end{equation}
after rearranging. Now, substituting this equation into itself recursively gives
\begin{align}
f_1(x)=&y(x)-R\Bigl[y(x+\delta)-Rf_1(x+2\delta)\Bigr]\\
           =&y(x)-Ry(x+\delta)+R^2f_1(x+2\delta) \\
           =&y(x)-Ry(x+\delta)+R^2\Bigl[y(x+2\delta)-Rf_1(x+3\delta)\Bigr]\\
           =&y(x)-Ry(x+\delta)+R^2y(x+2\delta)-R^3f_1(x+3\delta) \\
           =&y(x)-Ry(x+\delta)+R^2y(x+2\delta)-R^3\Bigl[y(x+3\delta) \\
            &-Rf_1(x+4\delta)\Bigr] \\
           =& \dotsc& \\
           =&y(x)-Ry(x+\delta)+R^2y(x+2\delta)-R^3y(x+3\delta) \\
            &+R^4y(x+4\delta)-R^5y(x+5\delta)-\dotsc.
\end{align}
This can be expressed concisely as
\begin{equation}
f_1(x)=\sum_{m=0}^{\infty}(-1)^mR^my(x+m\delta),
\end{equation}
which yields Eqn.~\eqref{extraction} after comparing the right-hand side with Eqn.~\eqref{transform_defn}.

\subsection{Linearity of the transform}
\label{Appendix:linearity}
A pleasing aspect of the transformation is that it is linear, i.e. for a constant $C$ and 
arbitrary functions $g$ and $h$,
\begin{equation}
\mathcal{T}[Cg] =C\mathcal{T}[g]
\end{equation}
and
\begin{equation}
\mathcal{T}[g+h]=\mathcal{T}[g]+\mathcal{T}[h],
\end{equation}
where $\mathcal{T}[g]\equiv \tilde{g}(x)$ denotes the result of applying the transform to $g(x)$.
These equations can be derived straightforwardly from Eqn.~\eqref{transform_defn}, or, for
the case of finite order $M$, Eqn.~\eqref{transform_order}.

\subsection{Derivation of Eqn. \eqref{background_lin}}
\label{Appendix:background}
Consider Eqn.~\eqref{extraction_gen} with $\eta(x)=Ax+C$. From the definition of the transformation with
finite order $M$, Eqn.~\eqref{transform_order}, we have
\begin{align}\label{background_deriv}
\tilde{\eta}(x)=&\sum_{m=0}^{M}(-1)^mR^m\Bigl[A(x+m\delta)+C\Bigr] \\
       =&(Ax+C)\sum_{m=0}^{M}(-R)^m + A\delta\sum_{m=0}^{M}m(-R)^m.
\end{align}
This becomes Eqn.~\eqref{background_lin} after applying the well-known equation
\begin{equation}\label{geometric_series}
\sum_{m=0}^{M}r^m=\frac{1-r^{M+1}}{1-r},
\end{equation}
to the first term, and an analogous equation (obtained from differentiating the above equation)
\begin{equation}
\sum_{m=0}^{M}mr^m=\frac{r}{(1-r)^2}\Bigl[1-(M+1)r^M+Mr^{M+1}\Bigr]
\end{equation}
to the second term. Note that both these equations are only valid for $r\neq 1$. Hence in
deriving Eqn.~\eqref{background_lin} from Eqn.~\eqref{background_deriv} we have tacitly assumed
that $(-R)\neq 1$, which is inconsequential since we are only interested in $R>0$.

\subsection{Derivation of Eqns. \eqref{noise_mean} and \eqref{noise_var}}
\label{Appendix:noise}
Consider if $\eta(x)$ is a random variable corresponding to additive Gaussian noise with mean and variance 
$\langle\eta(x)\rangle=0$ and $\langle\eta(x)^2\rangle=\sigma^2$ for all $x$. Taking the expected value of
$\tilde{\eta}(x)$  yields
\begin{equation}
\langle\tilde{\eta}(x)\rangle = \sum_{m=0}^{M}(-1)^mR^m\langle\eta(x+m\delta)\rangle,
\end{equation}
where we have used Eqn.~\eqref{transform_order} and exploited the linearity of the expected value 
(i.e., $\langle(a+b)\rangle=\langle a\rangle+\langle b\rangle$). Eqn.~\eqref{noise_mean} follows from this since 
$\langle\eta(x)\rangle=0$ for all $x$.

Now, since $\langle\tilde{\eta}(x)\rangle=0$, the variance in $\tilde{\eta}(x)$
is given by $\langle\tilde{\eta}(x)^2\rangle$. Noting that
\begin{align}
\tilde{\eta}(x)^2=&\sum_{m,n=0}^{M}(-1)^mR^m\eta(x+m\delta)(-1)^nR^n\eta(x+n\delta) \\
               =&\sum_{m,n=0}^{M}(-1)^{m+n}R^{m+n}\eta(x+m\delta)\eta(x+n\delta), \\
\end{align}
it follows that
\begin{equation}
\langle\tilde{\eta}(x)^2\rangle
=\sum_{m,n=0}^{M}(-1)^{m+n}R^{m+n}\langle\eta(x+m\delta)\eta(x+n\delta)\rangle,
\end{equation}
where again we have exploited the linearity of the expected value.

To proceed further we assume that there are no correlations in the noise over length-scales $\delta$ in $x$, in
which case $\langle\eta(x+m\delta)\eta(x+n\delta)\rangle=0$ for $n\neq m$. With this, all terms $n\neq m$ in 
the above equation vanish, leaving
\begin{equation}
\langle\tilde{\eta}(x)^2\rangle
=\sum_{m=0}^{M}(-1)^{2m}R^{2m}\langle\eta(x+m\delta)^2\rangle
=\sum_{m=0}^{M}(-1)^{2m}R^{2m}\sigma^2,
\end{equation}
where in the last equality we have used the fact that $\langle\eta(x)^2\rangle=\sigma^2$ for all $x$. Noting that
$(-1)^{2m}=1$, this can in turn be re-expressed as
\begin{equation}
\langle\tilde{\eta}(x)^2\rangle=\sigma^2\sum_{m=0}^{M}(R^2)^m,
\end{equation}
which yields Eqn.~\eqref{noise_var} after using Eqn.~\eqref{geometric_series}, where note that Eqn.~\eqref{geometric_series}
is only valid if $r\neq 1$ and hence Eqn.~\eqref{noise_var} only applies if $R^2\neq 1$, or equivalently $R\neq \pm 1$.
For the special case $R=1$ the above equation can be easily shown to be equivalent to Eqn.~\eqref{noise_var_R_1}.

\subsection{Divergence in noise if $R=1$}
\label{Appendix:R1}
\begin{figure*}
    \includegraphics[width=\columnwidth]{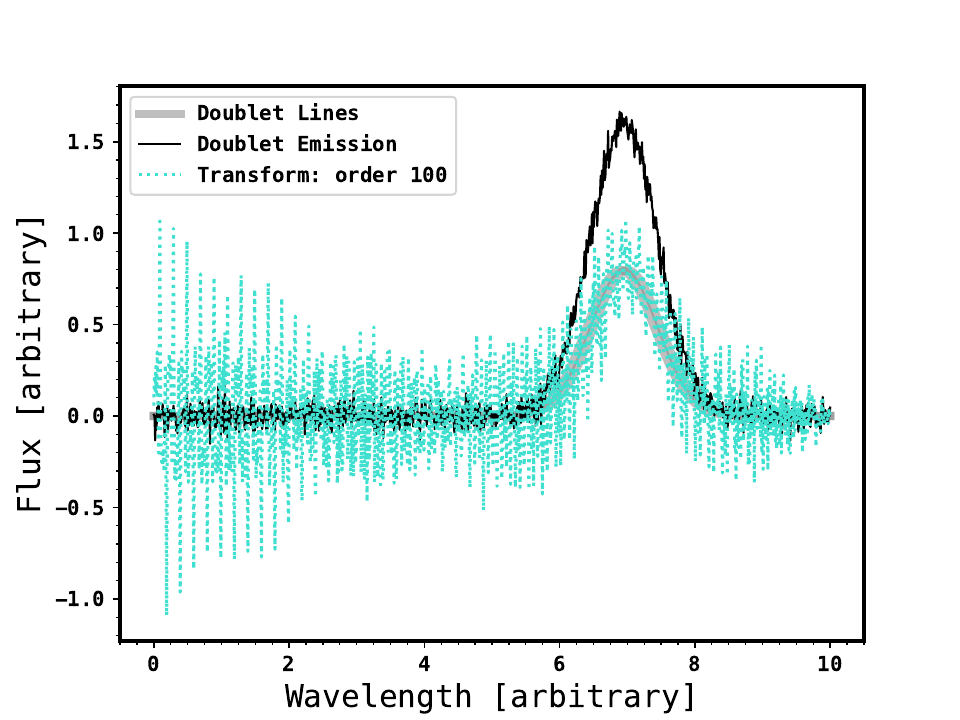}
    \includegraphics[width=\columnwidth]{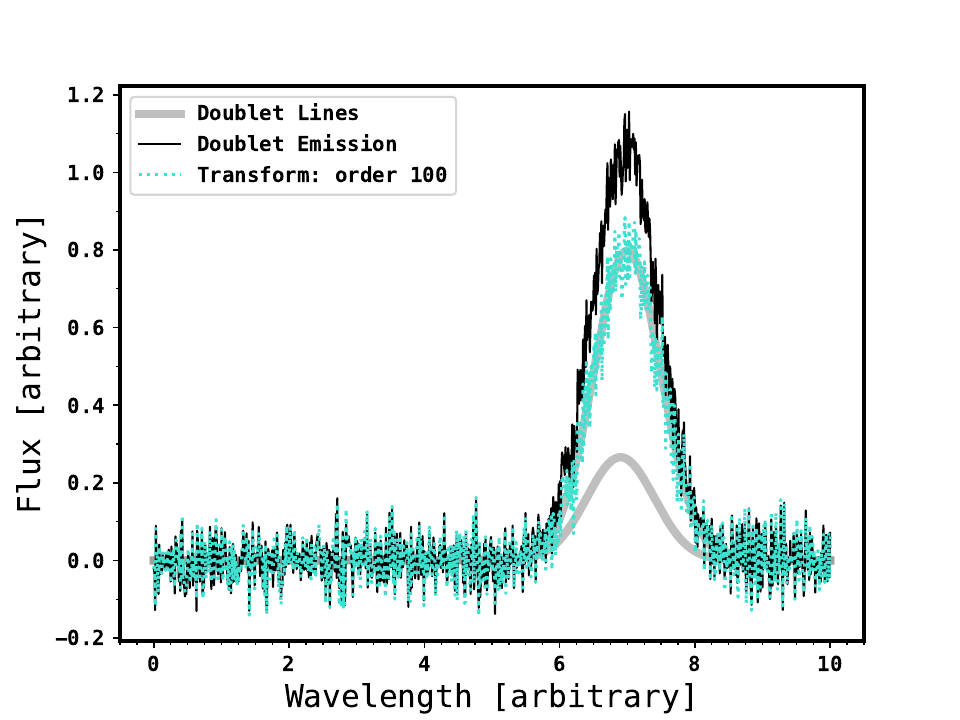}
    \caption{Simulations showing the effect of noise on a transform with a line ratio of $R=1$ for order=100. Left: A doublet comprised of two Gaussians with $R=1$, $\sigma=0.5$ and $\delta=0.1$. Divergence of the noise is noticeable over a large number of iterations in this case. The right panel shows the same noise level and line separation, but with a line ratio of $R=1/3$. The noise does not diverge in this case.}
    \label{fig:simu_r1noise}
\end{figure*}

Fig.~\ref{fig:simu_r1noise} (left panel) shows the effect of the transform on a doublet with line ratio $R=1$ in the presence of noise.
It is clear that the noise is magnified with increasing order of the transformation, as discussed in Section~\ref{S:noise}. 
Analogous results for $R=1/3$ are shown in the right panel of Fig.~\ref{fig:simu_r1noise}.

The transform can still be used for line ratios $R=1$, but great care should be taken; the number of iterations applied should be 
carefully considered.

\subsection{Uncertainty in $R$ and $\delta$}
\label{Appendix:delta_R_uncertainty}

Here we present an expression which describes the effect of applying the transformation with incorrect values of $R$ and 
$\delta$. We use $R_{\text{t}}$ and $\delta_{\text{t}}$ to denote the `true' line ratio 
and peak separation for the doublet under consideration, and $R$ and $\delta$ to denote the line ratio and peak 
separation actually used in the transform. Assuming that $R$ and $\delta$ are `close' to the true values, i.e.
that $\Delta R\equiv (R-R_{\text{t}})$ and $\Delta\delta\equiv (\delta-\delta_{\text{t}})$ are small, the error in
the transform of $y(x)$, given by Eqns.~\eqref{I_defn} and~\eqref{same_shape_defn}, from the true result 
$\tilde{y}(x)=f_1(x)$ (Eqn.~\eqref{extraction}), is as follows:
\begin{equation}\label{uncertainty_R_delta}
\Delta \tilde{y}(x) = \Bigl[ \widetilde{f_1}(x;R_{\text{t}},\delta_{\text{t}})-f_1(x) \Bigr]\frac{\Delta R}{R_{\text{t}}}
+ \Bigl[ \widetilde{f_1'}(x;R_{\text{t}},\delta_{\text{t}})-f_1'(x) \Bigr]\Delta\delta,
\end{equation}
where $\tilde{g}(x;R,\delta)$ denotes the transform of $g(x)$ using $R$ and $\delta$; and $g'(u)$ denotes the value of 
the derivative of $g(x)$ with respect to $x$, evaluated at $x=u$. Note that in the above $\widetilde{f'}_1(x;R_{\text{t}})$ 
is the transform of the derivative of $f_1(x)$ (and not the derivative of the transform). Note also that this
equation applies for $M\to\infty$, i.e. $\tilde{g}(x;R,\delta)$ is defined as the right-hand side of 
Eqn.~\eqref{transform_defn}.

The derivation of this equation is as follows. Expanding $\tilde{y}(x)$ as a Taylor series about $R=R_{\text{t}}$ and 
$\delta=\delta_{\text{t}}$ gives
\begin{equation}\label{Delta_tilde_y}
\begin{split}
\Delta \tilde{y}(x) &\equiv \tilde{y}(x;R,\delta)-\tilde{y}(x;R_{\text{t}},\delta_{\text{t}}) \\
&= \frac{\partial \tilde{y}(x;R,\delta)}{\partial R}\Delta R  + \frac{\partial \tilde{y}(x;R,\delta)}{\partial \delta}\Delta\delta,
\end{split}
\end{equation}
where the derivatives in the above are evaluated at $(R,\delta)=(R_{\text{t}},\delta_{\text{t}})$, and we have ignored
quadratic and higher-order terms in $\Delta R$ and $\Delta\delta$. 
The task now is to evaluate derivatives.

We begin with $\partial \tilde{y}(x;R,\delta)/\partial R$. Taking the derivative of Eqn.~\eqref{transform_defn} with respect to
$R$ gives
\begin{equation}
\frac{\partial\tilde{y}(x;R,\delta)}{\partial R} = \sum_{m=0}^{\infty}(-1)^mmR^{m-1}y(x+m\delta),
\end{equation}
which becomes 
\begin{equation}\label{deriv_R_1}
\frac{\partial\tilde{y}(x;R,\delta)}{\partial R} = R_{\text{t}}^{-1}\sum_{m=0}^{\infty}(-1)^mmR_{\text{t}}^{m}\Bigl[f_1(x+m\delta_{\text{t}})+R_{\text{t}}f_1(x+(m+1)\delta_{\text{t}})\Bigr]
\end{equation}
after using Eqns.~\eqref{I_defn} and~\eqref{same_shape_defn}, setting $(R,\delta)=(R_{\text{t}},\delta_{\text{t}})$, and
factorising.
Now, by expanding the right-hand side into two summations, we obtain $\partial\tilde{y}(x;R,\delta)/\partial R = S_1 + S_2$,
where
\begin{equation}
S_1=R_{\text{t}}^{-1}\sum_{m=0}^{\infty}(-1)^mmR_{\text{t}}^{m}f_1(x+m\delta_{\text{t}})
\end{equation}
and
\begin{equation}
S_2=\sum_{m=0}^{\infty}(-1)^mmR_{\text{t}}^{m}f_1(x+(m+1)\delta_{\text{t}}).
\end{equation}
$S_2$ can be expressed as
\begin{equation}
S_2=R_{\text{t}}^{-1}\sum_{m=1}^{\infty}(-1)^m(m-1)R_{\text{t}}^{m-1}f_1(x+m\delta_{\text{t}}).
\end{equation}
Hence we obtain
\begin{equation}
\frac{\partial\tilde{y}(x;R,\delta)}{\partial R} = S_1+S_2 = R_{\text{t}}^{-1}\sum_{m=1}^{\infty}
(-1)^mR_{\text{t}}^{m}f_1(x+m\delta_{\text{t}})
\end{equation}
after combining the two sums and exploiting the fact that the $m=0$ term in $S_1$ is 0. Moving the $m=0$ term in
Eqn.~\eqref{transform_defn} to the left-hand side and applying the resulting equation to the above equation, we obtain
\begin{equation}\label{deriv_R_final}
\frac{\partial\tilde{y}(x;R,\delta)}{\partial R} = \frac{1}{R_{\text{t}}}
\Bigl[ \widetilde{f_1}(x;R_{\text{t}},\delta_{\text{t}})-f_1(x) \Bigr].
\end{equation}

We now turn to $\partial \tilde{y}(x;R,\delta)/\partial \delta$. Taking the derivative of 
Eqn.~\eqref{transform_defn} with respect to $\delta$ gives
\begin{equation}
\frac{\partial\tilde{y}(x;R,\delta)}{\partial \delta} = \sum_{m=0}^{\infty}(-1)^mmR^{m}y'(x+m\delta),
\end{equation}
where recall that $y'(u)$ denotes the value of $\partial y(x)/\partial x$ evaluated at $x=u$. This becomes
\begin{equation}
\frac{\partial\tilde{y}(x;R,\delta)}{\partial \delta} = \sum_{m=0}^{\infty}(-1)^mmR_{\text{t}}^{m}\Bigl[f_1'(x+m\delta_{\text{t}})+R_{\text{t}}f_1'(x+(m+1)\delta_{\text{t}})\Bigr]
\end{equation}
after using Eqns.~\eqref{I_defn} and~\eqref{same_shape_defn}, and setting $(R,\delta)=(R_{\text{t}},\delta_{\text{t}})$.
Note that the right-hand side of the above equation resembles Eqn.~\eqref{deriv_R_1}. The difference is that the above has
$f_1'(x)$ instead of $f_1(x)$, and is missing the factor $R_{\text{t}}^{-1}$ present in Eqn.~\eqref{deriv_R_1}.
With this in mind, following the same procedure we used above to derive Eqn.~\eqref{deriv_R_final} from Eqn.~\eqref{deriv_R_1}, 
we obtain
\begin{equation}\label{deriv_delta_final}
\frac{\partial\tilde{y}(x;R,\delta)}{\partial R} = \widetilde{f_1'}(x;R_{\text{t}},\delta_{\text{t}})-f_1'(x).
\end{equation}

Finally, Eqn.~\eqref{uncertainty_R_delta} results from substituting Eqns.~\eqref{deriv_R_final} and~\eqref{deriv_delta_final}
into Eqn.~\eqref{Delta_tilde_y}.

\bsp	
\label{lastpage}
\end{document}